\documentclass[12pt]{article}

\usepackage{amsmath,amsfonts,amssymb,verbatim,graphicx}

\newcommand{\bx}{\mathbf{x}}
\newcommand{\bu}{\mathbf{u}}
\newcommand{\bv}{\mathbf{v}}
\newcommand{\bw}{\mathbf{w}}
\newcommand{\bnabla}{\boldsymbol{\nabla}}
\newcommand{\rd}{{\text{\rm d}}}
\newcommand{\be}{\begin{equation}}
\newcommand{\ee}{\end{equation}}
\newcommand{\la}{\label}
\newcommand{\ba}{\begin{array}{c}}
\newcommand{\ea}{\end{array}}

\title{On the Statistical Properties of the $3$D Incompressible Navier-Stokes-Voigt Model}

\author{Boris Levant${}^1$ \quad F\'abio Ramos${}^1$
\quad Edriss S. Titi${}^{1,2}$ \vspace{1cm}\\
${}^1$ Department of Computer Science and Applied Mathematics \\
Weizmann Institute of Science  \\
Rehovot 76100, Israel\\
${}^2$ Department of Mathematics \\
and  Department of Mechanical and  Aerospace Engineering \\
University of California \\
Irvine, CA  92697-3875, USA}

\date{January 4, 2009}

\begin{document}

\maketitle

\begin{center} {\it In honor of Professor Andrew Majda in
his 60th birthday}
\end{center}


\noindent{\bf Abstract.} The Navier-Stokes-Voigt (NSV) model of
viscoelastic incompressible fluid has been recently proposed as a
regularization of the 3D Navier-Stokes equations for the purpose of
direct numerical simulations. In this work we investigate its
statistical properties by employing phenomenological heuristic
arguments, in combination with Sabra shell model simulations of the
analogue of the NSV model. For large values of the regularizing
parameter, compared to the Kolmogorov length scale, simulations
exhibit multiscaling inertial range, and the dissipation range
displaying low intermittency. These facts provide evidence that the
NSV regularization may reduce the stiffness of direct numerical
simulations of turbulent flows, with a small impact on the energy
containing scales.


\medskip

\section{Introduction}

In this work, we study the statistical properties of the three-dimensional Navier-Stokes-Voigt (NSV) (sometimes it is written as Navier-Stokes-Voight) equations, an incompressible viscoelastic model introduced by Oskolkov in \cite{oskolkov1}, and proposed in \cite{klt} and \cite{KLT} as a smooth regularization of the $3$D Navier-Stokes equations, for the purpose of Direct Numerical Simulations (DNS).

Throughout the work we consider the NSV model subject to periodic or no-slip boundary conditions, and driven by a given force field $\mathbf{f}$. The velocity vector field, $\bu(\bx,t)$, and the scalar kinematic pressure, $p(\bx,t)$, are governed by the system of equations
\be
\left \{
\ba
\partial_t (\bu - \alpha^2\Delta \bu) -\nu \Delta \bu +\bu\cdot\nabla \bu   + \nabla p = \mathbf{f},\quad \bx\in\Omega,\\
\nabla\cdot \bu = 0,\quad \bx\in\Omega,\\
\bu(\bx,0)=\bu_0(\bx),\quad \bx\in\Omega,\\
\bu(\bx,t)=0\quad
\bx\in\partial\Omega,\quad\text{or}\quad\bu(\bx,t)\;\text{is}\;\text{periodic};
\ea \right. \la{aueq} \ee in the smooth domain $\Omega\subset
\mathbb{R}^3$, in the case of no-slip Dirichlet boundary condition,
and with basic periodic domain $\Omega = [0, L]^3\subset
\mathbb{R}^3$, when equipped  with periodic boundary conditions.
Here, $\alpha\geq 0$ is a given length scale parameter, and $\nu>0$
is a given kinematic viscosity, such that $\alpha^2/\nu$ is the
relaxation time of the viscoelastic fluid.

It is easy to see from \eqref{aueq} that besides having the same steady state solutions as the Navier-Stokes equations (NSE), the NSV model satisfies formally  the same infinite time Reynolds averaged equations as those for the NSE, suggesting a strong link with the statistical properties of turbulent flows.

As it was observed above, the NSV model presents an extra length scale associated to the viscoelasticity, the parameter $\alpha$, besides the well known Kolmogorov length scale, $\eta$, (see Section $3$ for its definition), which is usually associated to the smallest scales of motion in turbulent flows. For large values of the parameter $\alpha$, compared to $\eta$, we observe two distinct regions associated to the inertial range of the energy spectrum for the NSV model. The first one obeying the celebrated Kolmogorov $ k^{-2/3}$ power law (with anomalous correction), followed by a second range of length scales, where energy condensates, and it is simply equipartitioned.

The second power-law, however, vanishes as $\alpha$ is decreased, restoring the usual Navier-Stokes inertial range regime. This claim is supported by the numerical investigation of the regularized Sabra shell model of turbulence, as well as by rigorous results reported in the companion paper, \cite{RT}, concerning the weak convergence of invariant measures of the NSV to strong stationary statistical solutions of the Navier-Stokes equations.

We also present simulations of the Sabra shell model, with the NSV regularization term, which is displaying strong damping of dissipation range intermittent effects as $\alpha$ is increased,
due to a slowdown of the energy transfer timescales, see Section $3$ for more details.  We suggest that by tuning this parameter, we may attenuate the strong velocity fluctuations related to the intermittent events, reducing thus the the stiffness of DNS of turbulent flows, with only a small effect on the energy containing scales.

Another main advantage of the NSV model compared to other
regularization and sub-grid scale models used in ocean dynamics,
like hyperviscosity \cite{Lions2} or $\alpha$-models \cite{cht, ctk,
chesktiti, FHT2}, is the fact that in the presence of physical
boundaries the NSV model does not require any additional artificial
boundary conditions which cause difficulties and possibly exhibit
non-physical behavior in applications, such as non-physical boundary
layer, see, e.g., \cite{lesieur}.

\section{Some facts about the NSV model}

The Navier-Stokes-Voigt (sometimes written as Navier-Stokes-Voight) model of viscoelastic incompressible fluid, \eqref{aueq}, was introduced by Oskolkov in \cite{oskolkov1}, and pointed out by O. Ladyzhenskaya as one of the reasonable modifications of the Navier-Stokes equations, see \cite{ladyosk}. In \cite{oskolkov1}, A. P. Oskolkov studied and proved its solvability in different functional spaces. We refrain from giving technical details about regularity issues here, but we remark that these equations behave like a damped hyperbolic system, see \cite{klt}. Therefore, solutions do not experience fast (instantaneous) smoothening of the initial data, as it is for parabolic systems like the Navier-Stokes equations.

This fact could prevent the NSV model from being a reasonable modification of the Navier-Stokes equations, however, since we are proposing it as a model for direct numerical simulations of turbulent flows in statistical equilibrium, i.e., after the solutions reach the global attractor, we are mainly interested in its long time behavior. Indeed,  it was proved in \cite{KLT}, that solutions in the global attractor are smooth, if the forcing field is  smooth enough, even for initial data satisfying only finite kinetic energy and finite enstrophy (i.e. bounded in the Sobolev $H^1$-norm). In particular, in \cite{KLT}, it is shown in the periodic case, that if the forcing field is analytic, then the global attractor consists of analytic functions. This result, in conjunction with results proved in \cite{RT}, proves that if the forcing field is smooth enough, then averaged structure functions, with respect to an invariant measure for the NSV flow, display exponential decaying tail. One of the main goals of this paper is also to present direct numerical simulations, of the Sabra shell model analogue of the NSV model, supporting this asymptotic smoothening behavior.

Let $\Omega = [0, L]^3$. We denote by $L^p(\Omega)$, for $1 \le p
\le \infty$, and $H^m(\Omega)$ -- the usual Lebesgue and Sobolev
spaces of the periodic functions on $\Omega$ respectively. The inner product in the spaces $L^2(\Omega)$, which will be of particular interest for us, is given by
\[
(\bu,\bv) = \int_\Omega \bu(\bx)\cdot\bv(\bx) \;\rd\bx,
\]
and its associated norm is defined
\[
\left|\bu\right|=(\bu,\bu)^{1/2}.
\]
Let $\mathcal{F}$ be the set of all vector trigonometric polynomials on
the periodic domain $\Omega$, and denote
\begin{equation*}
\mathcal{V} = \Big \{ \varphi\in \mathcal{F} \;:\; \nabla\cdot
\varphi = 0, \; \text{and} \; \int_\Omega \varphi(x) dx = 0 \Big \}.
\end{equation*}
We set $H$ and $V$ to be the closures of $\mathcal{V}$ in the
$L^2(\Omega)$ and $H^1(\Omega)$ topologies, respectively. We equip
the spaces $H$ and $V$ with the inner products $(\cdot, \cdot)$ and
$((\cdot, \cdot))$, and  with the corresponding  norms $\left| \cdot
\right|$ and $\left| \left|\cdot \right|\right|$, respectively.

The NSV model satisfies the following energy equation for every $t\in \left[0,\infty\right)$,
\begin{equation}
\frac{d}{dt}(\frac{1}{2} \left| \bu (\cdot, t) \right|^2 + \frac{\alpha^2}{2} \left| \nabla \bu (\cdot, t) \right|^2) = (\mathbf{f}, \bu(\cdot, t)) - \nu \left| \nabla \bu (\cdot, t) \right|^2.
\end{equation}
Therefore, a positive quadratic conserved quantity in the inviscid, $\nu=0$, unforced, $\mathbf{f}=0$, and periodic or no-slip setting, which was proved rigorously in \cite{ctk}, is
\begin{equation}
S_2^{\alpha} = \frac{1}{2} \left| \bu \right|^2 + \frac{\alpha^2}{2}
\left| \nabla \bu \right|^2,
\end{equation}
which we call the $\alpha$-energy. The quantity
\begin{equation}
S_2 = \frac{1}{2} \left| \bu \right|^2
\end{equation}
is the usual kinetic energy, and we remark that it is not conserved for the inviscid unforced NSV equations.

Another conserved quadratic quantity is the $\alpha$-helicity
\begin{equation}
\label{helicity}
\Lambda_{\alpha} = (\bu - \alpha^2 \Delta \bu, curl(\bu)).
\end{equation}

Because the kinetic energy is not a conserved quantity, the arguments used by Kraichnan in \cite{kraichnan} (see also \cite{foias97}) to study the turbulent cascade scenario cannot be employed directly to the kinetic energy. However, all the investigation will be carried out instead to the conserved $\alpha$-energy, $S_2^{\alpha}$, and conclusions will be further recovered for the kinetic energy, $S_2$. This strategy was also used in \cite{cht, ctk, chesktiti, fht, gkt, lktt, lkt} for studying various $\alpha$ subgrid scale models of turbulence.

We denote by $P_{\text{LH}} : L^2 \to H$ -- the Helmholtz-Leray
orthogonal projection operator, and by $A = - P_{\text{LH}} \Delta=
- \Delta$ -- the Stokes operator subject to the periodic boundary
conditions with domain $D(A) = (H^2(\Omega))^3\cap V$. The operator
$A^{-1}$ is a positive definite, self-adjoint, compact operator from
$H$ into $H$. Therefore, there exists  a complete orthonormal basis
of $H$ formed by eigenvectors, $\{\mathbf{w}_{j}\}_{j\ge 1}$ of $A$,
with associated eigenvalues satisfying $0< \lambda_{j} \rightarrow
\infty$, when $j \rightarrow \infty$ (see, e.g.,
\cite{constantinfoias, fmrt, temam} for details).

The term $B(\bu, \bv) = P_{\text{LH}}((\bu \cdot \bnabla) \bv)$ is a
bilinear form associated with the inertial term. Taking the inner
product in $L^2$, i.e. in $H$, of the bilinear form with a third
variable yields a trilinear form
\[
b(\bu,\bv,\bw) = (B(\bu,\bv),\bw).
\]
An important relation for
the trilinear form is the orthogonality property (see, e.g.,  \cite{constantinfoias, fmrt, temam})
\begin{equation}
b(\bu,\bv,\bv) = 0.
\end{equation}

Let us define the component $\bu_k$ of a vector field $\bu\in H$, for a wavenumber $k$, by
$$
\bu_k = P_k \bu := \sum_{\lambda_j=k^2}\hat{u}_j\mathbf{w}_j,
$$
where $\mathbf{w}_{j}$, $j\ge 1$, are the eigenvectors of the operator $A$ with the corresponding eigenvalues $\lambda_{j}$, $j\ge 1$. We also define the component $\bu_{k', k''}$ by
$$
\bu_{k', k''} = \sum_{k' \leq k < k''} \bu_k.
$$
Then, we can write the projected NSV equations
\be
\label{projected}
\frac{d}{dt}(\bu_{k', k''}+\alpha^2 A\bu_{k', k''})+\nu A\bu_{k', k''}+B(\bu,\bu)_{k', k''}=\mathbf{f}_{k', k''}.
\ee

Let us now obtain the $\alpha$-energy budget. Taking the inner product in the space $H$ of the equation (\ref{projected}) with $\bu_{k', k''}$, we obtain
\be
\begin{aligned}
\frac{1}{2} \frac{d}{dt} (\left| \bu_{k', k''} \right|^2 & + \alpha^2\left| \nabla \bu_{k', k''} \right|^2) + \nu \left| \nabla\bu_{k', k''} \right|^2 =
\\
& = b(\bu, \bu, \bu_{k', k''}) - b(\bu, \bu, \bu_{k', k''}) + (\mathbf{f},\bu_{k', k''}) =
\\
& = [e^{\alpha}_{k'} -  e^{\alpha}_{k''}] + (\mathbf{f}, \bu_{k', k''}),
\la{tbud3}
\end{aligned}
\ee
where
\begin{equation} \label{defe}
e^{\alpha}_k(\bu)=e_{k}^{\alpha,\rightarrow}(\bu)-e^{\alpha,\leftarrow}_{k}(\bu)
\end{equation}
is the net rate of $\alpha$-energy transfer at $k$, and
\[
e_{k}^{\alpha, \rightarrow}(\bu) = -(B(\bu_{k_1, k},\bu_{k_1, k}),\bu_{k,\infty})
\]
represents the net rate of $\alpha$ -- energy from the lower modes to the higher modes, while that
\[
e_{k}^{\alpha,\leftarrow}(\bu)=-(B(\bu_{k, \infty},\bu_{k, \infty}),\bu_{k_1, k})
\]
represents the net rate of $\alpha$ -- energy from the higher modes to the lower modes.

\section{Averaged Energy Budget}

In this section, we follow  \cite{foias97}, \cite{fht} and
\cite{fmrt} to investigate the energy distribution scale-by-scale
for the $3$D NSV equations. Let $\langle \cdot \rangle$ denote
average with respect to an invariant measure, $\mu^\alpha$, for the
NSV semigroup (such a measure is known to exist for the NSV, see
\cite{RT}).  For the Navier-Stokes equations, assuming that there
exists an extensive range of wavenumbers, where the viscous
dissipation does not play a significant role, one can show that the
energy simply cascades through these length scales, with rate equals
to the mean energy dissipation rate for the NSV,
$\epsilon_\alpha=\nu\langle\left|\nabla\bu\right|^2\rangle$. For the
NSV equations, a similar scenario holds for the $\alpha$-energy
\[
S_2^{\alpha} = \frac{1}{2} \left| \bu \right|^2 + \frac{\alpha^2}{2} \left| \nabla \bu \right|^2.
\]

\subsection{Energy distribution scale-by-scale}

In the course of investigation of the $S_2^\alpha$ scaling, we will follow methods previously used in \cite{cht, ctk, chesktiti, fht, lktt, lkt}. As it is usual in the studies of homogeneous turbulence, we will consider the forcing $\mathbf{f}$ with finite number of the eigenmodes, i.e.
\be
\mathbf{f}=\sum_{\underline{ k}\leq k\leq\bar{ k}}\mathbf{f}_ k.
\ee
Let $ k''\geq k'>\bar{ k}$. If we take averages in \eqref{tbud3} with respect to an invariant measure, $\mu^{\alpha}$, we obtain the following balance equation
\be
\nu \langle \left| \nabla \bu_{k', k''} \right|^2 \rangle = \langle e^{\alpha}_{k'}(\bu) \rangle - \langle e_{ k''}(\bu) \rangle
+ \langle(\mathbf{f}, \bu_{k', k''}) \rangle = \langle e^{\alpha}_{ k'}(\bu)\rangle- \langle e^{\alpha}_{ k''}(\bu)\rangle,
\la{tbud5}
\ee
where $e^{\alpha}_{k}(\bu)$ was defined in (\ref{defe}).

The expression on the right-hand side of the last equality is the mean net $\alpha$-energy transfer in the energy shell $[k', k'']$. In particular, if we choose $ k''=\infty$, we obtain
\be
\nu \langle \left| \nabla \bu_{k, \infty} \right|^2 \rangle = \langle e^{\alpha}_{k}(\bu)\rangle.
\la{infitransf}
\ee
This expression shows that the net $\alpha$-energy transfer is positive for every $ k>\bar{ k}$.  Moreover, if we assume that there exists a range of wavenumbers, $\left[k', k'' \right]$,  where the left-hand side of \eqref{tbud5}, $\nu \langle \left| \nabla \bu_{k', k''} \right|^2 \rangle$, is very small, then the $\alpha$-energy transfer is near constant within this range, i.e.,
\be
\langle e^{\alpha}_{ k'}(\bu) \rangle \sim \langle e^{\alpha}_{ k''}(\bu) \rangle.
\ee
Defining $k_\tau = (\langle \left| \nabla \bu \right|^2 \rangle / \langle \left| \bu \right|^2 \rangle)^{1/2}$,
we can follow \cite{fmrt} to derive
\[
0 \leq 1 - \frac{\langle e^{\alpha}_{ k''}(\bu) \rangle}{\langle
e^{\alpha}_{ k'}(\bu) \rangle} \leq \left(\frac{k''}{k_\tau}
\right)^2 \left(1 - \left(\frac{k'}{k_\tau} \right)^2\right)^{-1}.
\]
Indeed,
\be
\begin{aligned}
1 - \frac{\langle e^{\alpha}_{ k''}(\bu) \rangle}{\langle
e^{\alpha}_{ k'}(\bu) \rangle} &= \frac{\langle \sum_{k'\le k <
k''}k^2 \left|\bu_k \right|^2 \rangle}{\langle \sum_{k'\le k} k^2
\left| \bu_k \right|^2 \rangle} \le
 \frac{ (k'')^2 \langle \sum_{k'\le k < k''} \left|\bu_k \right|^2 \rangle}{\langle
\sum_{k'\le k} k^2 \left| \bu_k \right|^2 \rangle}\\ \nonumber & \le
\frac{ (k'')^2 \langle \sum_{ k } \left| \bu_k \right|^2
\rangle}{\langle \sum_{k'\le k} k^2 \left| \bu_k \right|^2 \rangle}
= \left(\frac{k''}{k_\tau} \right)^2\frac{ \langle \sum_{ k } k^2
\left| \bu_k \right|^2 \rangle}{\langle \sum_{k'\le k} k^2 \left|
\bu_k \right|^2 \rangle} \\ \nonumber & \le \left(\frac{k''}{k_\tau}
\right)^2\left(1- \frac{ \langle \sum_{ k \le k' } k^2 \left| \bu_k
\right|^2 \rangle}{\langle \sum_{ k} k^2 \left| \bu_k \right|^2
\rangle}\right)^{-1} \\ \nonumber & \le \left(\frac{k''}{k_\tau}
\right)^2 \left(1 - \left(\frac{k'}{k_\tau} \right)^2\right)^{-1}.
\end{aligned}
\ee
Therefore, if $k'' \ll k_\tau$, then $\langle e^{\alpha}_{
k''}(\bu) \rangle \sim \langle e^{\alpha}_{ k'}(\bu) \rangle$, which
means that there is no leak of energy in this range. Of course, we
cannot expect this condition to be fulfilled for every forcing term
$\mathbf{f}$. In the last section, where we investigate numerically
the NSV spectrum scenario, we provide Sabra shell model simulations
satisfying an analogue condition, assuring the cascade scenario.

We denote by $\epsilon_\alpha$, the total energy dissipation rate for the NSV,
\[
\epsilon_{\alpha} = \nu \langle| \nabla \bu|^2 \rangle.
\]
Now, we want to investigate the distribution of the inviscid conserved quantity, the $\alpha$-energy, scale-by-scale. We define the following characteristic velocities at scale $ k$:
\[
\mathbf{U}^{(0)}_{ k} = \langle |\bu_{ k}|^2 \rangle^{1/2},
\]
and
\[
\mathbf{U}^{(\alpha)}_{ k}=(1+\alpha^2 k^2)\langle |\bu_{ k}|^2\rangle^{1/2}.
\]
We denote the characteristic $\alpha$-energy at scale $ k$ by
\be
S_2^{\alpha}( k)=\frac{1}{2}\mathbf{U}^{(0)}_{ k}\mathbf{U}^{(\alpha)}_{ k},
\la{s2def}
\ee
 The characteristic kinetic energy at scale $ k$ by
\[
S_2(k) = \frac{1}{2} (\mathbf{U}^{(0)}_{k})^2.
\]
With this notation, we can write the $\alpha$-energy as
\[
S_2^{\alpha} = \sum_{ k} S_2^{\alpha}(k) = \frac{1}{2} \sum_{k} \mathbf{U}^{(0)}_{ k} \mathbf{U}^{(\alpha)}_{k}.
\]
and the kinetic energy as
\[
S_{2} = \sum_{k} S_{2}(k) = \frac{1}{2} \sum_{k} (\mathbf{U}^{(0)}_{ k})^{2}.
\]
In the inertial range, the $\alpha$-energy transfer time-scale can be defined as
\be
\la{pitransf}
t^{\mbox{transf}}_{ k}=\frac{S_2^{\alpha}( k)}{\epsilon_{\alpha}}.
\ee
Following arguments used by Kraichnan in \cite{kraichnan}, (see also \cite{foias97}, \cite{fht} and \cite{rosa}), in the inertial range, the eddies of size $ k^{-1}$, in average, transfer their characteristic $\alpha$-energy to neighboring eddies in the time, $t^{\mbox{transf}}_{ k}$, it takes to travel their own length, $ k^{-1}$, i.e.,
\be
t^{\mbox{transf}}_{ k} = \frac{1}{k U_{ k}},
\la{timetransf}
\ee
where $U_k$ is the characteristic velocity at scale $ k$. Substituting \eqref{timetransf} in \eqref{pitransf}, and setting $\alpha=0$, we recover the $ k^{-2/3}$ scaling for the inertial range of the $S_2$ structure function, that was theoreticaly predicted for the Navier-Stokes equations by Kolmogorov in \cite{kolmogorov} (we remark that the $\kappa^{-2/3}$ scaling for the structure function is commonly quoted in terms of its  correspondent energy spectrum density, which obeys a $\kappa^{-5/3}$ power law). This sort of argument also leads to the double cascading scenario for $2$D turbulence described in \cite{kraichnan} (see also \cite{foias97}).

For the NSV case, the situation is complicated by the fact that we have two different characteristic velocities, $U^{(0)}_{ k}$ and $U^{(\alpha)}_{ k}$. And, in fact, any log-convex combinations of them would give us a possible characteristic velocity, see, e.g., \cite{cht, ctk, chesktiti, ILT}.

Real world turbulent flows, however, present anomalous scaling, i.e., structure functions deviate significantly from the Kolmogorov predictions in \cite{kolmogorov}, see, e.g., \cite{frisch, LP98, she}.  This anomalous behavior is present in some phenomenological models, as the Sabra shell models introduced in \cite{LP98}. For example, in \cite{LP98}, the scaling computed for the $S_2$ structure function was $0.72$, slightly deviating
from the Kolmogorov $-2/3$ scaling. The nature of inertial range intermittency is a topic of current intense research in the turbulence community, see, e.g., \cite{falkovich} for more details.

\begin{figure}
\label{davinci2}
\begin{center}
\includegraphics[width=1\textwidth]{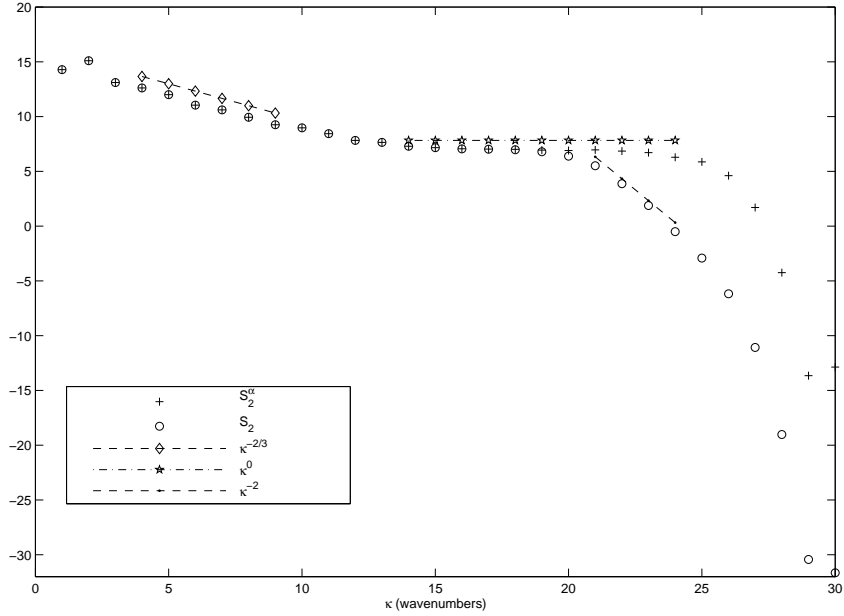}
\caption{Log-log plot of Sabra-NSV simulation with parameters $\nu=10^{-9}$, and $\alpha=10^{-5}$. $(+)$ $S_2^{\alpha}( k)$ - Characteristic $\alpha$-Energy at scale $ k$. $(\circ)$ $S_2( k)$ - Characteristic Kinetic Energy at scale $ k$.}
\end{center}
\end{figure}

\begin{figure}
\label{davinci}
\begin{center}
\includegraphics[width=1\textwidth]{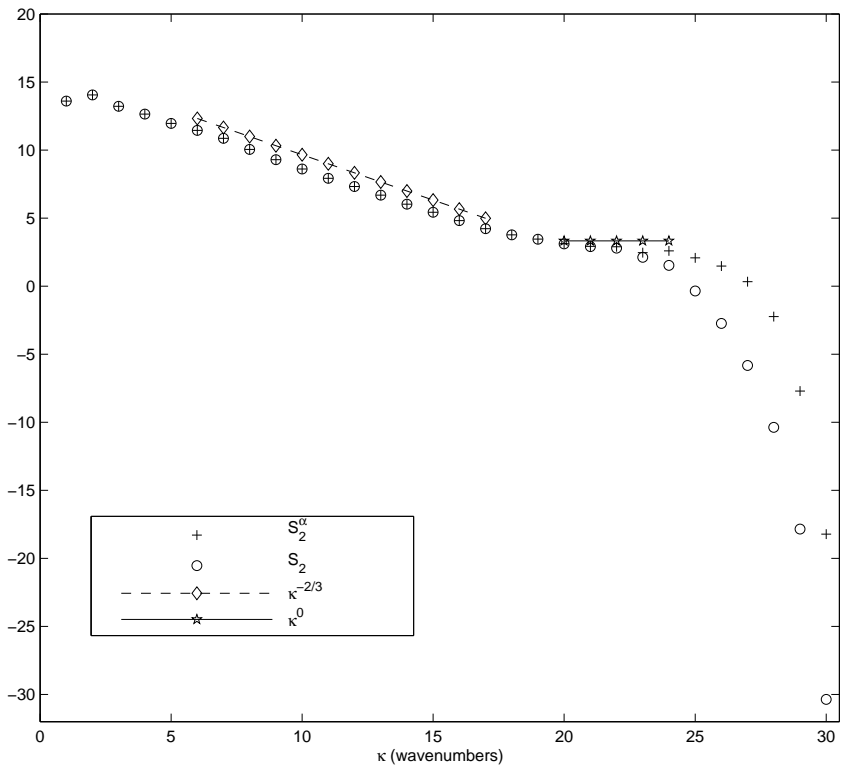}
\caption{Log-log plot of the Sabra-NSV simulation with parameters $\nu=10^{-9}$, and $\alpha=10^{-6}$. $(+)$ $S_2^{\alpha}( k)$ - Characteristic $\alpha$-Energy at scale $ k$. $(\circ)$ $S_2( k)$ - Characteristic Kinetic Energy at scale $ k$.}
\end{center}
\end{figure}

Simulations of the Sabra-NSV shell model, presented in the coming section, clearly display two distinct power-laws for $S_2^{\alpha}$. For large values of $\alpha$ compared to the Kolmogorov dissipation length scale, $ \eta := (\nu^3 / \epsilon_\alpha)^{1/4}$, we observe a range with scaling slightly deviating from the $ k^{-2/3}$ Kolmogorov scaling (see Section~\ref{sabra-nsv-section} for more details), and another range with a nearly power zero scaling, see
Figures~\ref{davinci2} and~\ref{davinci}. This distribution can be explained if we set up the transfer time-scale, $t_{ k}^{\mbox{transf}}$, in equation \eqref{timetransf}, as a function of the translational velocity, $U_ k\sim
 U_ k^{(0)}$:
\be
t_{ k}^{\mbox{transf}}=\frac{1}{ k U_ k^{(0)}}\sim \frac{(1+\alpha^2 k^2)^{1/2}}{ k (U_{ k}^{(0)})^{1/2}(U_{ k}^{(\alpha)})^{1/2}}.
\ee
In fact, substituting the expression above into \eqref{pitransf}, we obtain
\be
S_2^{\alpha}(k) \sim \epsilon_\alpha^{2/3} k^{-2/3} (1 + \alpha^2 k^2)^{1/3}.
\la{s2shape}
\ee
And, therefore, for  $ k\ll\alpha^{-1}$, we have a $ k^{-2/3}$ range, while that for $\alpha\approx k^{-1}$, we have a power zero range, just as it is observed in the shell model simulations in the next section.

We remark that this scenario of two power laws in the inertial range was first proposed in \cite{fht} for the NS-$\alpha$ model, and then for the rest of the $\alpha$ models in \cite{cht, ctk, chesktiti, ILT, lktt, lkt}.

\subsection{Smallest scales of motion}

The idea of a smallest scale of motion in turbulent flows was introduced by Kolmogorov in \cite{kolmogorov}. This can be obtained by simple dimensional analysis, but can also be obtained by comparing the energy transfer time scale, $t_ k^{\mbox{transf}}$ and the dissipative timescale, $t_ k^{\mbox{dissip}}:=\frac{1}{\nu k^2}$, leading to the Kolmogorov length scale $\eta = \left(\nu^3 / \epsilon_\alpha \right)^{1/4}$ (see, e.g., \cite{frisch}).

\begin{figure}
\label{davinci3}
\begin{center}
\includegraphics[width=0.78\textwidth]{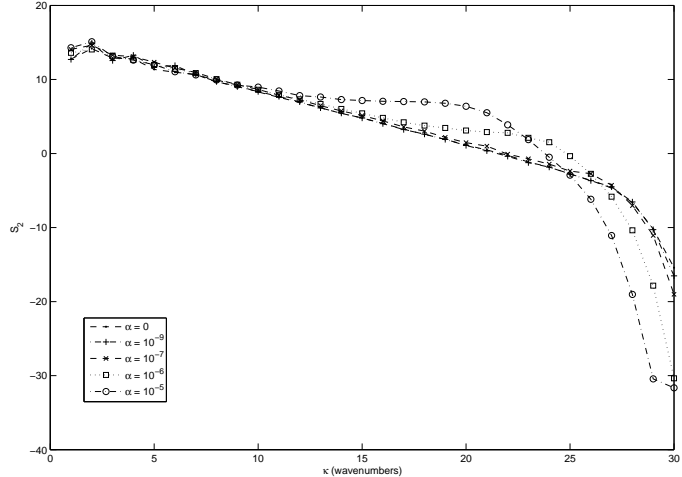}
\caption{Log-log plot of the characteristic kinetic energy scale-by-scale for the Sabra-NSV simulations with $\nu=10^{-9}$, and different values of the parameter $\alpha$.}
\end{center}
\end{figure}

In the same spirit, for the NSV case, we define the smallest scale of motion, $\eta^{NSV}$, as the scale where energy transfer time scale, $t_ k^{\mbox{transf}}$, equals the dissipative time scale, $t_ k^{\mbox{dissip}}$. In order to estimate it, we first obtain explicitly the transfer time scale. Comparing \eqref{s2def} with \eqref{s2shape}, we obtain the following expression for the characteristic velocity at scale $ k$:
\be
U^{(0)}_ k\sim \epsilon_\alpha^{1/3} k^{-1/3}(1+\alpha^2 k^2)^{-1/3}.
\ee
Therefore,
\be
t_{ k}^{\mbox{transf}}=\frac{1}{ k U^{(0)}_ k}=\frac{(1+\alpha^2 k^2)^{1/3}}{\epsilon_\alpha^{1/3} k^{2/3}}.
\la{transfshape}
\ee
Thus, when we equate \eqref{transfshape} with $t_ k^{\mbox{dissip}}$, we see that for $\alpha<\eta$, the smallest scale of motion, $\eta^{NSV}$, is exactly the Kolmogorov length scale, i.e., $\eta^{NSV}=\eta$. However, when $\alpha>\eta$, it is easy to see that
\be
\frac{\eta^{NSV}}{\eta}\sim \left(\frac{\alpha}{\eta}\right)^{1/3}.
\la{reduced}
\ee
Equation \eqref{reduced} shows that if we choose $\eta\ll\alpha$, the degrees of freedom of the NSV flow are significantly reduced in comparison to the NSE flow, which is also a great advantage
concerning direct numerical simulations. Figure \ref{davinci3}
supports this fact for Sabra-NSV simulations.

\subsection{Comparison with previous works}

There is no consensus about the correct transfer time scale in previous works. In the first studies on the subject, for example in \cite{fht}, in the context of the NS-$\alpha$ model, it was argued that the time scale, $t_ k^{\mbox{transf}}$, should be setup by the \textit{translational velocity}, i.e., $\mathbf{U}_ k\sim\mathbf{U}^{(0)}_ k$. This argument was reinforced by new large scale simulations by Graham et al. in \cite{mininni}, although the numerical results are mostly inconclusive, due to a still low resolution.

Recently, Lunasin et al. in \cite{lkt} and \cite{lktt}, argued that the time-scale should be set up by the combination of velocities appearing at the conservation law. This leads to the observed power laws for DNS of the $2$D Leray-$\alpha$, in \cite{lkt}, and of the $2$D NS-$\alpha$ model, in \cite{lktt}.

So far, there is no clear phenomenological explanation for the scaling here described for $S_2^{\alpha}$. It is in consonance with the arguments for the $3$D NS-$\alpha$, obtained in \cite{fht}, where DNS simulations are still inconclusive, but in dissonance with the arguments of the $2$D simulations in \cite{lktt, lkt}, where high resolution DNS were employed, and clear power laws were presented. It is reasonable that the form of the nonlinear term, responsible for the energy transfer mechanism, should play a role in the transfer time scale of the conserved quantity. However, none of the former arguments contemplate it, and a robust explanation for the observed scaling is still lacking.

We also observe that in the shell model simulations in the next section, the distribution of kinetic energy scale by scale, $S_2$, presents an inertial range with three distinct power laws. Indeed, the log-log plot of $S_2$ in
Figure \ref{davinci} follows the shape of $S_2^\alpha$,
with its two power laws, until it reaches wavenumbers
obeying a third power law, with scaling $ k^{-3}$, just
before the dissipation range takes place. We remark that
this intermediate range, with a power zero spectrum, which
is reminiscent from the $\alpha$-energy spectrum, has never been
theoretically predicted or numerically computed in any of the
former works about $\alpha$ models, see, e.g., \cite{cht, ctk, chesktiti, fht, ILT, lktt, lkt}.

\section{Sabra-NSV shell model simulations} \label{sabra-nsv-section}

Due to the extraordinary complexity of the hydrodynamic equations, many simplified models, based on the phenomenological theories of turbulence, have been proposed in order to investigate the statistical scenario of complex flows. In this section, we use a modification of the Sabra shell model of turbulence, which describes the evolution of complex Fourier-like components of a scalar velocity field denoted by $u_n\in \mathbb{C}$. The associated one-dimensional wavenumbers are denoted by
$k_n$, where the discrete index $n$ is referred to as the ``shell
index''. The equations of motion of the Sabra shell model of
turbulence were introduced in \cite{LP98}, and they have the following form

\begin{equation}
\label{eq_Sabra}
\frac{d u_n}{d t} = i (a k_{n+1} u_{n+2}
u_{n+1}^* + b k_{n} u_{n+1} u_{n-1}^* - c k_{n-1} u_{n-1} u_{n-2}) -
\nu k_n^2 u_n + f_n,
\end{equation}
for $n = 1, 2, 3, \dots$, with the boundary conditions $u_{-1}= u_{0} = 0$. The wave numbers $k_n$ are taken to be
\begin{equation} \label{eq_freq}
k_n = k_0 \lambda^n,
\end{equation}
with $\lambda  > 1$ being the shell spacing parameter, and $k_0
> 0$. Although the equation does not capture any geometry, the scale  $L = k_0^{-1}$ is frequently considered as a fixed typical length scale of the
model. In an analogy to the Navier-Stokes equations $\nu > 0$
represents a kinematic viscosity and $f_n$ are the Fourier
components of the forcing.

We consider the following regularization of the model associated to the NSV equations, which we  denominate as the Sabra-NSV shell model
\begin{equation} \label{eq_Sabrareg}
\begin{aligned}
\frac{d u_n}{d t} =& \frac{i}{1+\alpha^2 k_n^2} (a k_{n+1} u_{n+2} u_{n+1}^* + b k_{n}
u_{n+1} u_{n-1}^* - c k_{n-1} u_{n-1} u_{n-2})\\
& - \frac{\nu k_n^2}{1+\alpha^2 k_n^2} u_n +
\frac{f_n}{1+\alpha^2 k_n^2},
\end{aligned}
\end{equation}
for $n = 1, 2, 3, \dots$, for fixed $\alpha > 0$.

The three parameters of the model $a, b$ and $c$ are real. Following \cite{LP98}, we require that in the inviscid ($\nu = 0$) and unforced
($f_n = 0$, for all $n$) case, the model should have at least one formal
quadratic positive definite quantity to be invariant. Such a
quantity will represent the $\alpha$-energy in the system. Indeed,
in order to require that the $\alpha$-energy
\begin{equation}
\label{eq_energy}
\mathbb{E} = \sum_{n=1}^\infty (1+\alpha^2 k_n^2)|{u_n}|^2,
\end{equation}
will be formally conserved, we assume the following relation between the
parameters of the model, which we will refer as an ``energy
conservation assumption''
\begin{equation} \label{eq_energy_cons_assumption}
a + b + c = 0.
\end{equation}

Moreover, in the inviscid and unforced case the model possesses
another formal quadratic invariant
\begin{equation} \label{eq_helicity}
\mathbb{W} = \sum_{n=1}^\infty \bigg( \frac{a}{c} \bigg)^n
(1+\alpha^2 k_n^2)|u_n|^2.
\end{equation}
For $\frac{a}{c} < 0$ this quantity is not sign definite and thus it
is common to associate it with the ``helicity'' -- in an analogy to
the three-dimensional turbulence, \eqref{helicity}.

\noindent{\bf{Remark.}} For $\alpha=0$, that is, when we are in the pure Sabra regime, the above mentioned conservation laws are only formal, due to a possible lack of regularity of solutions of the inviscid Sabra shell model. However, by following \cite{ctk} (see also \cite{CLT_Sabra}), we can prove, when $\alpha>0$, global existence and uniqueness for the inviscid Sabra-NSV shell model, and therefore, it can be proved that $\mathbb{E}$ and $\mathbb{W}$ are rigorously conserved in the inviscid and unforced case.

Without lost of generality we may assume that $k_0 = 1$. Next, by rescaling the time
\[
t \to a t,
\]
and using the ``energy conservation assumption''
(\ref{eq_energy_cons_assumption}) we may set
\begin{equation}\label{eq_sabra_parameters}
a = 1, \;\;\; b = -\theta, \;\;\; c = \theta - 1.
\end{equation}
Therefore, the Sabra shell model is in fact a three-parameter family
of equations with parameters $\nu > 0$, $\theta$, and $\lambda$.
We are interested in the case where the shell sizes grow
geometrically (see (\ref{eq_freq})), therefore we limit ourselves to
$\lambda > 1$.

The three-dimensional parameters regime corresponds to $0 < \theta <
1$, as $\mathbb{W}$ is not sign definite (see, e.g., \cite{CLT_Sabra}, \cite{LP98}). In that regime we can rewrite relation (\ref{eq_helicity}) in
the form
\begin{equation} \label{eq_helicity2}
\mathbb{W} = \sum_{n=1}^\infty (-1)^n k_n^\beta
(1+\alpha^2 k_n^2)|u_n|^2,
\end{equation}
for
\begin{equation}\label{eq_alpha}
\beta = - \log_\lambda |\theta - 1|.
\end{equation}

In our simulations, we set $\theta=1/2$, $\lambda=2$, $k_0=1$, and with constant forcing, of order one, in both real and imaginary parts of the second and third modes.

We first comment on the results derived in the last section for the characteristic energy distribution scale by scale. In shell models, we refer to these quantities as the second order structure function, and define them by
\[
S_2(k_n)=\langle\left| u_n\right|^{2}\rangle,
\]
and
\[
S_2^\alpha(k_n)=\langle\left| u_n\right|^{2}+\alpha^2 k^2\left| u_n\right|^{2}\rangle.
\]
Figure $1$ and Figure $2$  show the results of  simulations for $\nu=10^{-9}$, and for $\alpha=10^{-5}$, and $\alpha=10^{-6}$, respectively. They show two clear distinct power-laws, one with a scaling of $k_n^{-0.66}$, and another one with a constant scaling.
Figure $3$ shows the effect of varying the parameter $\alpha$ for fixed $\nu=10^{-9}$. As expected, the secondary power law becomes less prominent as $\alpha$ is decreased.

Now, we comment on some issues about intermittency. One of the main characteristics of turbulent fluid flows, that is present in the Sabra shell model, is its dissipation-range intermittency. This is characterized by violent fluctuations of very short duration in the energy dissipation rate,
\[
\epsilon_\alpha = \langle \sum_{n} k_{n}^2 \left| u_{n} \right|^2\rangle.
\]
Small velocity fluctuations in high wavenumbers play a key role in this phenomenon, see, e.g. \cite{frisch}, and the rise of such fluctuations imposes severe challenges to the direct numerical simulations of turbulent flows. Therefore, the attenuation of its effects finds many applications.

Looking carefully at the equation \eqref{eq_Sabrareg}, we observe that as $k_n$ becomes large, the term $1 + \alpha^2 k_n^2$ in the denominator will damp the velocity fluctuations. Therefore, the energy dissipative intermittency must be significantly attenuated for large values of the relaxation time parameter. Figures $4$, $5$ and $6$ present the energy dissipation fluctuation signal, $\epsilon'_\alpha / \epsilon_\alpha = (\sum_{n\ge 1} k_n^2 \left| u_{n}(t)\right|^2)/\langle\sum_{n\ge 1} k_n^2\left| u_{n}(t)\right|^2\rangle$, for different settings. The intermittency becomes strongly attenuated as we increase the length parameter $\alpha$. This fact might find real world applications, and worth further investigations.

Another main characteristic of the Sabra model, as observed in \cite{LP98}, is that it exhibits an inertial range, with moments of the velocity depending on $ k_n$ as power laws with nontrivial exponents, i.e., $\langle\left| u_n\right|^{q}\rangle \sim k_n^{-\xi_{q}}$, where $\xi_q$ depends nonlinearly on $q$. From now on, we refer to these moments as structure functions. For even $q=2m$, we use the usual definition
\[
S_{2m}(k_n)=\langle\left| u_n\right|^{2m}\rangle.
\]
For odd $q=2m+1$, we use the following definition
\[
S_{2m+1}(k_n)=Im\langle u_{n-1} u_n u_{n+1}\left| u\right|^{2m-1}\rangle.
\]
For small values of the length parameter $\alpha$, i.e., for $\alpha \ll \eta$, where $\eta$ is the Kolmogorov length scale, we do not observe any significant deviations for the exponents of the structure functions from the pure Sabra model, with $\alpha=0$. In Figure $7$, we show some structure functions for a simulation with $\nu=10^{-9}$, and $\alpha=10^{-7}$.  The values of the anomalous exponents are listed in the tigure.

For larger values of the length parameter $\alpha$ compared to $k_{n_d}$, the analysis of the anomalous exponents is complicated by the presence of the double power law, and since we are concerned with the small relaxation time regularization, we will not present any detailed analysis of it here.

However, we want to announce an interesting fact that we plan to investigate in a forthcoming work. Self-similarity is a crucial hypothesis in K$41$ theory, see, e.g., \cite{frisch}. Intermitent dynamics in turbulence, however, is inconsistent with the self-similarity hypothesis, leading to several modifications of this hypothesis, see, e.g., \cite{frisch}, for more details.  One can measure the departure from self-similarity in the inertial range by looking at the flatness of the signal in this range, see \cite{frisch}. The flatness is defined as $F( k)=\langle S_4( k)\rangle/\langle S_2( k)^2\rangle$, and a departure from a constant value means a lack of self-similarity. For turbulent-fluid flows, like in shell models, this term deviates significantly from a constant as $ k = k_n$ increases, even within the inertial range.

The nature of inertial range intermittency in shell models is not well understood. The influence of high wavenumbers strong velocity fluctuations might play a role, just like in real turbulent flows, as suggested by Landau in his famous criticisms of $K41$ theory, see, e.g., \cite{frisch}. Because for the NSV model, we have strong damping of high wavenumbers velocity fluctuations, we calculated the flatness of the Sabra NSV model to investigate the relationship between the two kinds of intermittency.

We observe that the inertial range intermittency is significantly reduced as the parameter $\alpha$ becomes large when compared to the Kolmogorov dissipation length scale, $\eta$. For example, for the pure Sabra model with $\nu=10^{-9}$, we observe a flatness of the order $F(k_n) \sim k_n^{0.14}$, while that for the Sabra-NSV with same viscosity, and values of $\alpha=10^{-6}$, and $\alpha=10^{-5}$, the flatness is, respectively, $F(k_n)\sim k_n^{0.09}$, and $F(k_n)\sim k_n^{0.05}$.

For $\nu=10^{-8}$, the departure was from $F(k_n)\sim k_n^{0.14}$, for the pure Sabra model, to  $F(k_n)\sim k_n^{0.09}$, for $\alpha=10^{-5}$, and to $F(k_n)\sim k_n^{0.002}$ for $\alpha=10^{-4}$. Another interesting remark is that we fit the first power law, and we observed a trend towards the $K41$ scaling with less and less anomaly, as the relaxation time was increased.

Genuine fluctuations for odd order structure functions, see \cite{LP98}, prevented us from performing more detailed analysis of this pattern, and we are currently investigating it by using a special stochastic forcing that eliminates this complication. We expect to report the results of this investigation soon.

\begin{figure}
\begin{center}
\includegraphics[width=0.7\textwidth]{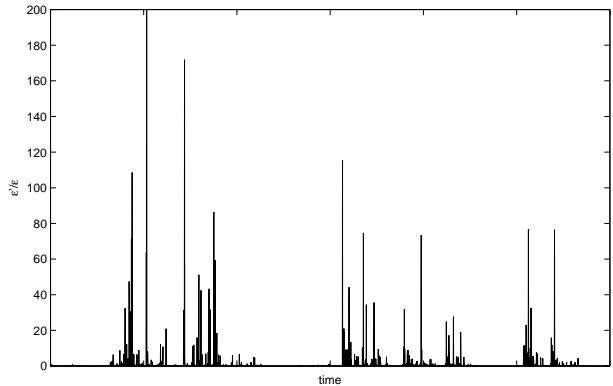}
\caption{Log-log plot of the energy dissipation fluctuation $\epsilon'_\alpha / \epsilon_\alpha$. Parameters: $\nu=10^{-9}$. $\alpha=0$.}
\end{center}
\end{figure}

\begin{figure}
\begin{center}
\includegraphics[width=0.7\textwidth]{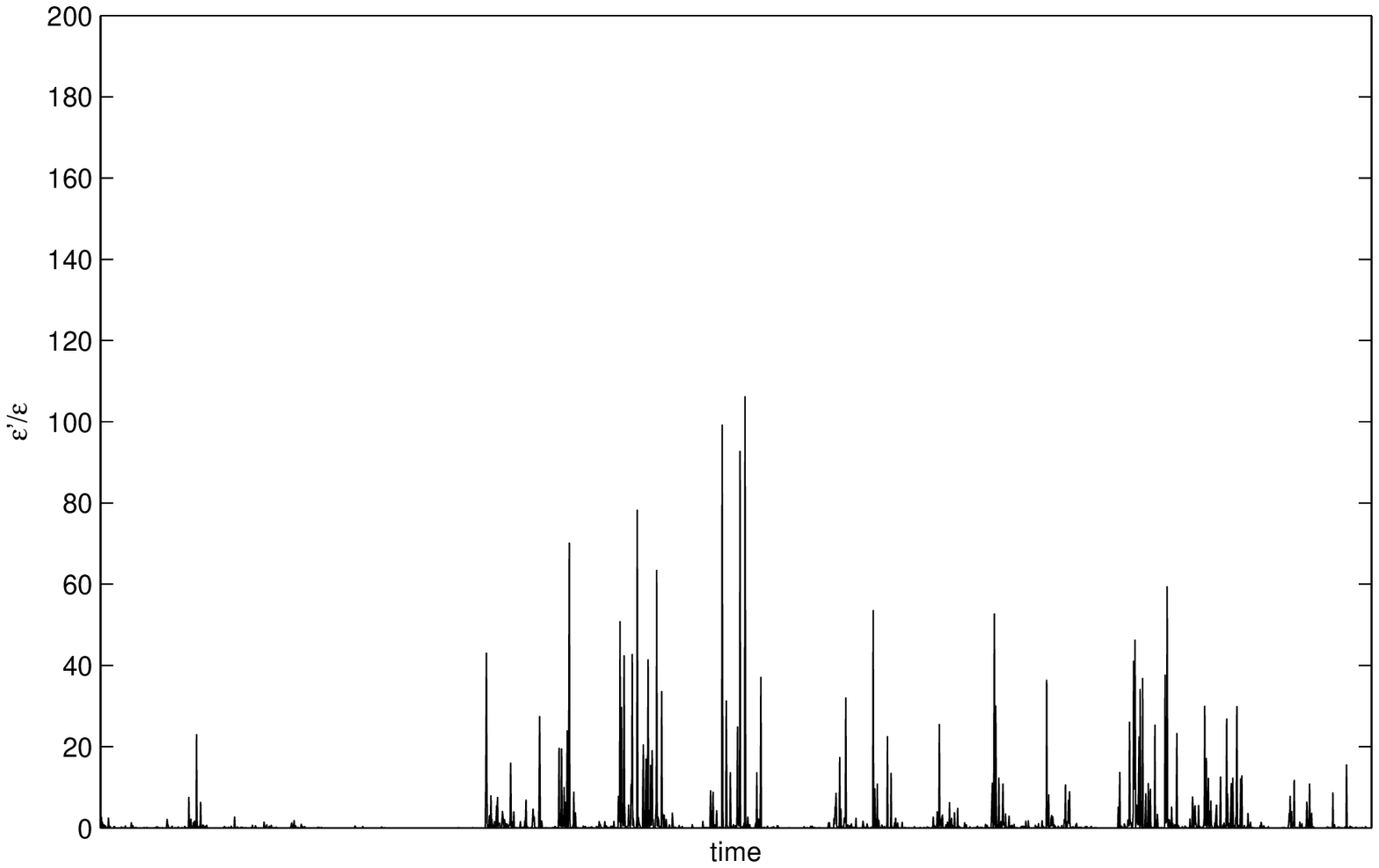}
\caption{Log-log plot of the energy dissipation fluctuation $\epsilon'_\alpha / \epsilon_\alpha$. Parameters: $\nu=10^{-9}$. $\alpha=10^{-7}$.}
\end{center}
\end{figure}

\begin{figure}
\begin{center}
\includegraphics[width=0.7\textwidth]{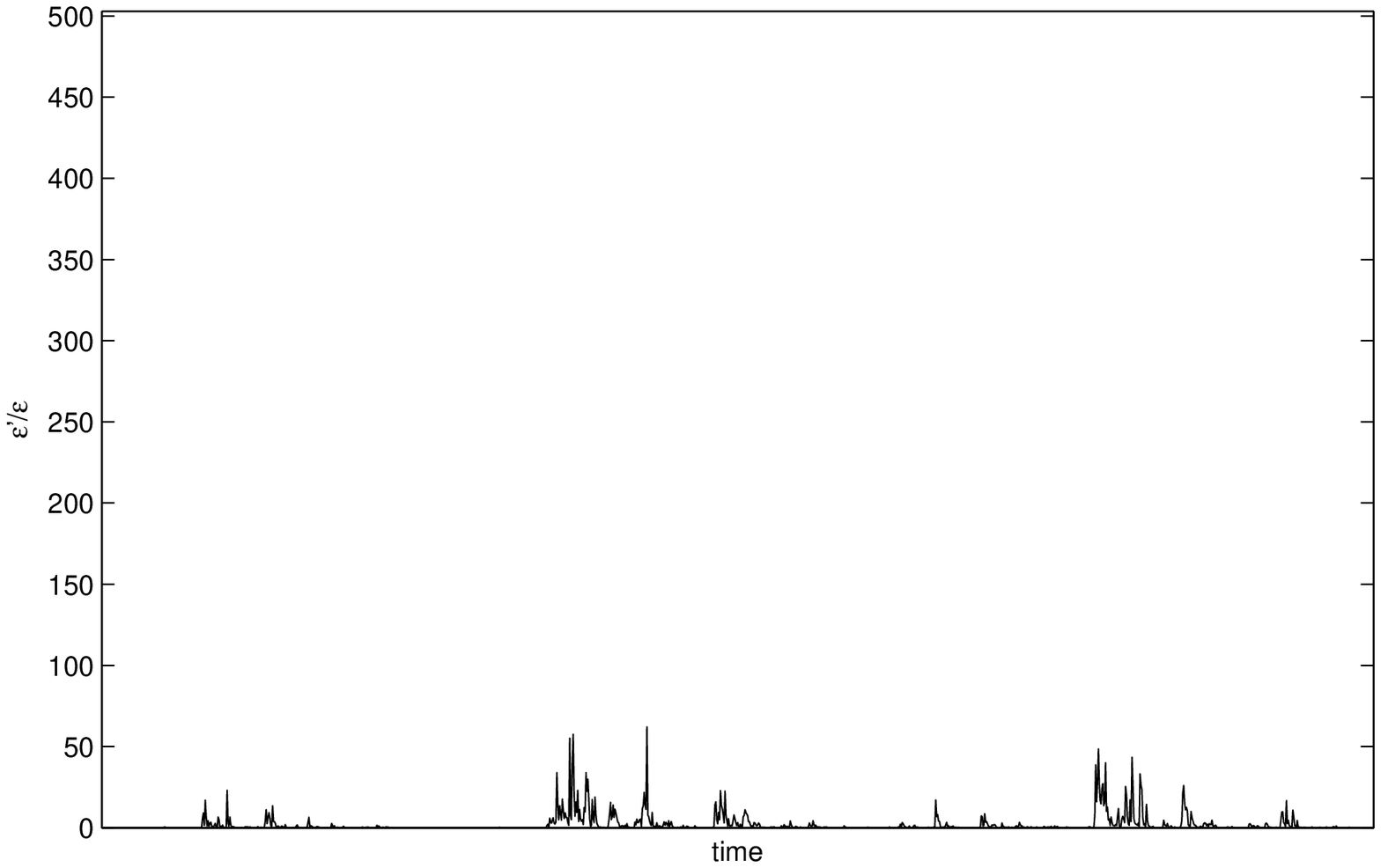}
\caption{Log-log plot of the energy dissipation fluctuation $\epsilon'_\alpha / \epsilon_\alpha$. Parameters: $\nu=10^{-9}$. $\alpha=10^{-6}$.}
\end{center}
\end{figure}

\begin{figure}
\begin{center}
\includegraphics[width=1\textwidth]{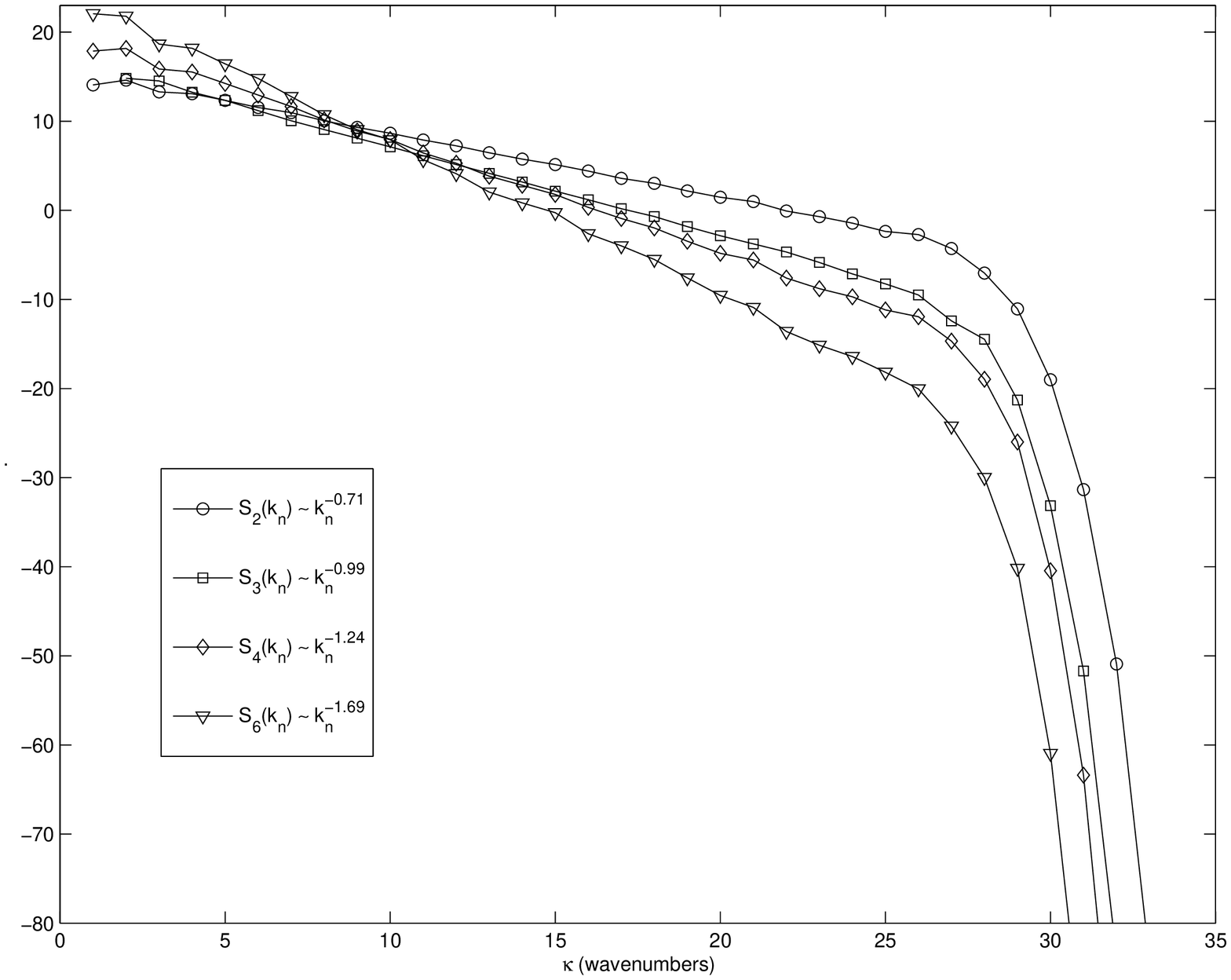}
\caption{Log-log plot of the structure functions, $S_2$, $S_3$, $S_4$ and $S_6$ for a simulation with $\nu=10^{-9}$, and $\alpha=10^{-7}$. One can also find the corresponding scaling of the structure functions in the inertial range.}
\end{center}
\end{figure}

\vspace{1cm}
\noindent{\bf{Acknowledgment.}}\,\,\, The authors would like to thank Professor Itamar Procaccia for the stimulating discussions. This work was partly supported by the NSF grants no. DMS-0504619 and no. DMS-0709228, the ISF grant no. 126/2, and the BSF grant no. 2004271. F.R. was also supported by the Koshland Center for Basic Research at the Weizmann Institute of Science. F.R. also wants to thank Pablo Mininni and Susan Kurien for fruitful discussions while this work was in progress.

\end{document}